\documentstyle[epsf,cite,12pt,a4]{article}
\newcommand{\be}{\begin{equation}}
\newcommand{\ba}{\begin{eqnarray}}
\newcommand{\ee}{\end{equation}}
\newcommand{\ea}{\end{eqnarray}}

\newcommand{\GeV}{\;\mbox{GeV}}
\newcommand{\eV}{\;\mbox{eV}}
\newcommand{\cm}{\;\mbox{cm}}
\newcommand{\mb}{\;\mbox{mb}}

\newcommand{\secs}{\;\mbox{s}}
\newcommand{\simlt}{\stackrel{<}{{}_\sim}}
\newcommand{\simgt}{\stackrel{>}{{}_\sim}}

\newcounter{currequation}
\newenvironment{subeqns}{\setcounter{currequation}{\value{equation}}%
\stepcounter{currequation}\setcounter{equation}{0}%
\begin{eqnarray}}{\end{eqnarray}{\setcounter{equation}{\value{currequation}}}%
}
\begin{document}
\title{Long-range interactions between dark-matter particles in a model with a cosmological,
spontaneously-broken chiral symmetry
}
\author{Saul Barshay and Georg Kreyerhoff\\
III.~Physikalisches Institut A\\
RWTH Aachen\\
D-52056 Aachen}
\maketitle
\begin{abstract}
In a cosmological model with a chiral symmetry, there are two, dynamically-related spin-zero fields,
a scalar $\phi$ and a pseudoscalar $b$. These fields have self-interactions. Spontaneous symmetry breaking
results in a very massive scalar particle with $m_\phi\cong 5 \times 10^{11}\GeV$, and a nearly massless, (Goldstone-like)
pseudoscalar particle with $0< m_b \simlt 2.7\times 10^{-6}\eV$. One or both particles can be part of dark matter.
There are coherent long-range interactions (at range $\sim 1/m_b \simgt 10\cm$), from exchange of a $b$ particle 
between a pair of $b$ particles, a pair of $\phi$ particles, and between a $\phi$ and a $b$. We compare the strength
of potentials for the different pairs to the corresponding gravitational potentials (within the same range $\sim 1/m_b$), and
show that the new force dominates between a b pair, that gravitation dominates between a $\phi$ pair, and that the
potentials are comparable for a $\phi$-$b$ pair. The new interaction strength between a $b$ pair is comparable
to the gravitational interaction between a $\phi$ pair; its possibly greater coherent effect originates in the
possibility that the number density of a very light $b$ can be greater than that of a massive $\phi$.
We consider these results in the context of recent speculations concerning possible effects of special
forces between dark-matter particles on certain galactic, and inter-galactic, properties.
\end{abstract}
In this note, we point out the strength, and certain possible physical consequences, of coherent long-range interactions
which can exist only between dark-matter particles. These interactions arise naturally in a model with a 
cosmological, spontaneously-broken chiral symmetry \cite{ref1,ref2}. There have been a number of speculations
about such forces. For example, one motivation \cite{ref3} has been to try to find a physical role for 
massless fields (dilatons) which exist in certain hypothetical, popular theories. Another recent speculation
has been motivated \cite{ref4} more directly, by certain empirical facts concerning galactic, and inter-galactic
structure, which may conflict with the results of simulations carried out in the standard cosmological model
($\Lambda$CDM). As we point out below, results of our present investigation based upon a specific dynamical
model, have  marked analogies to these phenomenological considerations. \cite{ref4}\footnote{
We thank Prof.~P.~J.~E.~Peebles for a communication about his concern for these questions.
}.

In the chiral field theory \cite{ref5,ref6}, there is a self-interaction term $\lambda(\phi^2+b^2)^2$. There is
a spontaneous symmetry breaking at a very high energy scale $\phi_c\simlt M_P \cong 10^{19}\GeV$, the
Planck mass. We have identified $\phi$ with a scalar inflaton field. \cite{ref1,ref2}
Identification
with the inflaton allows us to fix a dimensionless parameter $\lambda  \cong 3\times 10^{-14}$, 
essentially
empirically, \cite{ref7,ref8} from the observed CMB fluctuations. The definite, unusual 
result is that one possible dark-matter particle is at a very high mass, $m_\phi \sim 2\sqrt{2}\sqrt{\lambda}
\phi_c \sim 5\times 10^{11}\GeV$, for $\phi_c\sim 10^{18}\GeV$.\cite{ref2} This is also an implicit result of the phenomenology in ref.~4, where a 
classical scalar, (dark-energy) field with present magnitude of the order of $M_P$ is coupled linearly to a
fermionic, dark-matter field (unless the dimensionless coupling parameter, denoted by $y$ there \cite{ref4}, is
miniscule).

A new dynamical element which we introduced \cite{ref1}, is a further spontaneous symmetry breaking through 
a non-zero vacuum expectation value for the pseudoscalar $b$ field, $F_b\cong 5.5\eV$. Thus, CP invariance is 
spontaneously broken. The low energy scale is independently fixed by a coupling of the $b$ field to a neutrino
(in particular, to the heaviest ordinary neutrino, presumably $\nu_\tau$). This neutrino then acquires a mass
of $g_\nu F_b \cong 0.05\eV$, for a typical $g_\nu \sim 10^{-2}$ \cite{ref2}. Two further results follow from
the model.
\begin{itemize}
\item[(1)] There is a natural possibility for generation of a significant ($\simgt 10^{-9}$) antineutrino-neutrino
asymmetry in the early universe, as a consequence of CP noninvariance \cite{ref1}.
\item[(2)] There is a residual vacuum energy density with a magnitude of $\lambda F_b^4 \cong 2.7\times 10^{-47}\GeV^4$,
which can be identified with an effective cosmological constant \cite{ref1}. (See Appendix.)
\end{itemize}

As a result of $F_b\neq 0$, the self-interaction acquires trilinear coupling terms, $4\lambda F_b \tilde{b}^3$ and
$4\lambda F_b\tilde{\phi}^2 \tilde{b}$, where we denote the particle quanta of the $\phi$ and $b$ fields
by $\tilde{\phi}$ and $\tilde{b}$.\footnote{
These interaction terms in the Lagrangian density are formally odd under CP. We assume that a CP even coupling
that would lead to rapid decay of $\phi$ to a $b$-pair, is absent. (This is an explicit symmetry breaking by deletion,
although explicit breaking is usually associated with addition of special terms.). Combinatorial factors arising
from identical fields at the vertices are not included in the illustrative estimates below, nor are contact
interactions.
} 
These terms give rise to coherent long-range interactions, as illustrated
in Fig.~1 for a pair of $b$ particles. Below we give the effective, attractive (Yukawa) potentials between a $b$ pair,
a $\phi$-$b$ pair, and a $\phi$ pair, where the interacting particles are assumed to be nearly static i.~e.~with a
very small, intrinsic relative velocity.
\begin{subeqns}
b-b & & -\frac{(4\lambda F_b)^2}{4m_b^2}\frac{e^{-m_b r}}{4\pi r} = - \left(\frac{\lambda}{8\pi}\right)\frac{e^{-m_b r}}{r}\\
\phi-b & &  -\frac{(4\lambda F_b)^2}{4m_b m_\phi}\frac{e^{-m_b r}}{4\pi r} = - \left(\frac{\lambda}{8\pi}\right)
\left(\frac{F_b}{\phi_c}\right)\frac{e^{-m_b r}}{r}\\
\phi-\phi & &  -\frac{(4\lambda F_b)^2}{4m_\phi^2}\frac{e^{-m_b r}}{4\pi r} = - \left(\frac{\lambda}{8\pi}\right)
\left(\frac{F_b}{\phi_c}\right)^2 \frac{e^{-m_b r}}{r}
\end{subeqns}
At distances within $1/m_b$, with the exponential factor of order unity, we compare these attractive potentials 
(with $\phi_c \sim 0.1 M_P$ \cite{ref2}, $F_b\sim 5.5\eV$ \cite{ref1}, $m_b\cong 2 \sqrt{2}\sqrt{\lambda}F_b$) 
to the corresponding potentials from gravity:
\begin{subeqns}
b-b && -G \frac{m_b^2}{r} = - 8 \lambda \left(\frac{F_b}{M_P}\right)^2 \frac{1}{r}\\
\phi-b && -G \frac{m_\phi m_b}{r} = - 8 \lambda \left(\frac{F_b\phi_c}{M_P^2}\right) \frac{1}{r}\\
\phi-\phi &&  -G \frac{m^2_\phi}{r} = - 8 \lambda \left(\frac{\phi_c}{M_P}\right)^2 \frac{1}{r}
\end{subeqns}
We have written the gravitational constant $G\cong 1/M_P^2$. Clearly, for $b$-$b$ the effective coupling strength in (1a),
$\lambda/8\pi$, is much greater than the $8\lambda (F_b/M_P)^2$ in (2a). The strengths in (1b) and (2b) are comparable.
For $\phi$-$\phi$, the gravitational strength in (2c) is much greater than that in (1c). An important comparison
of strengths is between the new potential between a pair of very light, dark-matter $b$ particles in (1a), and the
gravitational potential between a pair of very heavy, dark-matter particles in (2c). The ratio of (1a) to (2c) is 
independent of the coupling parameter $\lambda$; it is simply $1/64\pi (\phi_c/M_P)^2 \cong 1 $. This is much 
like the result of the phenomenology in ref.~4; Eq.~(18) there gives a ratio  of a ``fifth'' force in the dark
sector to the gravitational force, as $1/4\pi G\phi^2$, where $\phi$ is a (time-varying) dark-energy field
with present magnitude $\simlt M_P$. 

For a mass $m_b \cong 2\sqrt{2}\sqrt{\lambda}F_b = 2.7\times 10^{-6}\eV$, the range for the interaction of a pair
of $b$ particles is $\sim 1/m_b \sim 10\cm$. The cross section for elastic scattering of an isolated pair,
with very small, intrinsic relative velocity, is simply $\sigma = \lambda^2/16\pi m_b^2 \sim 1\mb$ (up to approximately
$1\;{\mathrm{b}}$ with enhancement from the square of a combinatorial factor). Note the compensation
of the small factor $\lambda^2$ by the small mass in the dimensional factor $1/m_b^2$. To possibly have effects upon structure
within a galactic dimension, the number density of such light particles must be very large.\footnote{
For our hypothetical, very massive dark-matter particles $\phi$, a characteristic galactic number density is
$\sim 10^{-12} \cm^{-3}$. For a significant contribution to the total dark-matter energy density, a $b$-particle
number density of the order of $10^{26}$ larger is necessary ($\sim 10^{14}\cm^{-3}$). Note that this large
factor is approximately the ratio of dynamical times $(1/F_b)/(1/\phi_c)$, and that both $F_b$ and $\phi_c$ are
determined by dynamical considerations other than the size of contributions to the dark-matter energy density. \cite{ref1,ref2}
Such large number densities have been considered for a zero-momentum condensate of QCD axions, in which case they
depend upon the square of a high energy scale. It is possible that a present large number density of $b$ particles
could originate in production by a time-varying gravitational field \cite{ref8} over an extended time interval
during the early rapid expansion of the universe, say from $\sim 10^{-36}\secs$ \cite{ref1} (where the expansion
parameter $H(t)$ is of order of $10^{11}\GeV$, equivalent to the momentum of $\sim 10^{26}$ coherently-produced
$b$ i.~e.~with coherent momenta $\sim 10^{-6}\eV \sim 1/R$, and $R$ the dimension at that time of the observable
universe). Small primary, dark-matter fluctuations can also occur at a relatively late time, where $H(t)$ is $\simgt F_b$.
Then, $\delta\rho/\rho \sim H|\delta b|/|\dot{b}| \simgt m_b/F_b \sim 3\sqrt{\lambda}$, for $|\delta b|\sim m_b$,
and assuming that $|\dot{b}|\sim F_b^2$ can be relevant to an early epoch.
}
One likely effect is the formation of very massive dark-matter cores at the very center of many galaxies \cite{ref9,ref10}.
At larger distances from the center, other possible structure effects have been the subject of conflicting claims, made
under certain assumptions concerning collision cross sections and the motion of the dark-matter particles
\cite{ref11,ref12,ref13,ref14}. It is noteworthy that if there was relativistic motion of $b$ with momentum
$p> m_b$ in the earliest epoch of structure formation, a relevant $\sigma$ reduced by $(m_b^2/p^2)^2 \simgt (10^{-4})^2$ still
allows for a mean free path then less than 1 pc.

We consider the possibility that the $b$ mass is much smaller\footnote{
For example, $F_b$ may be at a (metastable) maximum of the effective potential for the $b$ field, where the
second derivative (the effective, squared $b$ mass), is dynamically brought to near zero by some mechanism.
}, but greater than the present Hubble parameter, and
that the $b$ particles are essentially a zero-momentum condensate, with  (galactic) number density similar to that for
the massive $\phi$. If the potential from Fig.~1 extends to distances between galaxy groups, then the coherence factor
for all $b$ particles is similar to (possibly even greater than) that for all $\phi$ particles. From the potentials 
in (1a) and (2c), the coherent effect of the new long-range force between $b$ dark matter is comparable to 
gravitation between $\phi$ dark matter (which dark matter gives the dominant contribution to the total, near-critical
energy density).

In summary, the idea discussed here is that it is possible that dark matter has both a very massive and a very light
particle content. With spontaneous breaking of CP invariance, the self-interaction of these dark-matter bosons can naturally
generate attractive, coherent long-range potentials\footnote{In the model of ref.~1, a long-range interaction between
a pair of slowly-moving (dark-matter) neutrinos is also noted.}, which can augment the gravitational attraction
between concentrations of large numbers of dark-matter particles.\footnote{A  remark of similar nature is made in section 5
of ref.~4. There  also a second kind of dark-matter particle is considered, which is (initially) light; it is assumed
to have relativistic motion. } 

Again, we thank Prof.~P.~J.~E.~Peebles for his encouraging communications.

\section*{Appendix}
In the model with a cosmological, chiral symmetry \cite{ref1}, the magnitude of the residual
vacuum energy density is estimated to be $|\rho_\Lambda|\sim |-\lambda F_b^4| \sim 2.7 \times 10^{-47}\GeV^4$,
with $\lambda$ and $F_b$ determined independently from the measured CMB fluctuations and from (a largest)
ordinary neutrino mass, respectively. In order to have a positive effective cosmological constant, it is
possible to consider an additional ``gravitational'' contribution of the form \cite{ref15,ref16} $(-1/48) F_{\mu\nu\sigma\rho}
F^{\mu\nu\sigma\rho}$ where $\sqrt{-g}F_{\mu\nu\sigma\rho} = C\epsilon_{\mu\nu\sigma\rho}$, as introduced
in the Lagrangian density by Duff and van Nieuwenhuizen \cite{ref15}. The essential new hypothesis here, 
is that the scale of the constant $C=(m_b F_b/\sqrt{2})$; then the contribution to the vacuum energy density
is $(1/4) (m_bF_b)^2 = 2\lambda F_b^4$. The total vacuum energy density equivalent to an effective cosmological
constant is $+\lambda F_b^4$. It is assumed that the additional contribution occurs in time only after the
minimum of the effective potential is reached at $b=F_b$, and so a stable, non-zero mass $m_b = 2\sqrt{2}\sqrt{\lambda}F_b$.
(The additional contribution is not negative as in \cite{ref16}, where $F_{\mu\nu\sigma\rho}$ is taken as purely
imaginary because it is used to cancel an assumed positive vacuum energy density of unknown origin.)
There is the possibility here of a unification of dark energy and dark matter, since $b$ particles
may constitute the latter.

\newpage
\begin{figure}[t]
\begin{center}
\mbox{\epsfysize 13cm \epsffile{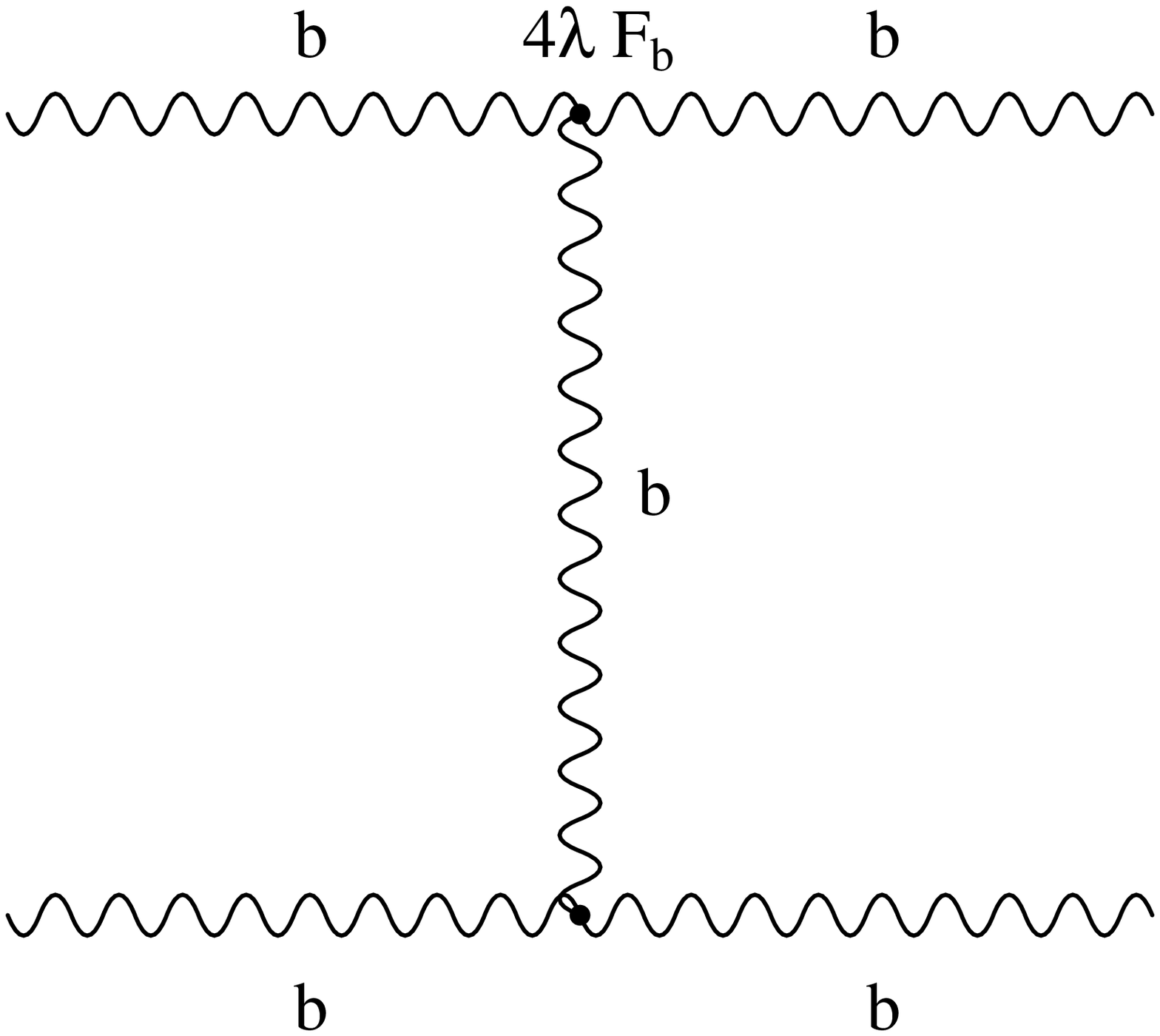}}
\end{center}
\caption{
Exchange of a very light $b$ particle generates a long-range potential between a pair of $b$ (with assumed
very small, intrinsic relative velocity). The vertex strength in the chiral model with spontaneously-broken
CP invariance $(F_b\neq 0)$, is shown.
}
\end{figure}

\begin{thebibliography}{9}
\bibitem{ref1} S.~Barshay and G.~Kreyerhoff, Mod.~Phys.~Lett. {\bf A19} 2899 (2004). In refs.~13, the
first page number should read 167, and the third page number should read 369.
\bibitem{ref2} S.~Barshay and G.~Kreyerhoff, Eur.~Phys.~J.~{\bf C5}, 369 (1998)\\
	       S.~Barshay and G.~Kreyerhoff, Z.~Phys.~{\bf C75}, 167 (1997); Erratum: Z.~Phys.~{\bf C76}, 577 (1997)
\bibitem{ref3} T.~Damour, G.~W.~Gibbons amd C.~Gundlach, Phys.~Rev.~Lett.~{\bf 64}, 123 (1990)
\bibitem{ref4} G.~R.~Farrar and P.~J.~E.~Peebles, astro-ph/0307316v2
\bibitem{ref5} M.~Gell-Mann and M.~Levy, Nuovo Cimento {\bf 16}, 705 (1960)
\bibitem{ref6}  B.~W.~Lee, {\it Chiral Dynamics}, Gordon and Breach Science Publishers, Inc., 1972
\bibitem{ref7} H.~Kurki-Suonio and G.~J.~Mathews, Phys.~Rev.~{\bf D50}, 5431 (1994)
\bibitem{ref8} P.~J.~E.~Peebles and A.~Vilenkin, Phys.~Rev.~{\bf D59}, 063505 (1999)
\bibitem{ref9}  S.~Barshay and G.~Kreyerhoff, Nuovo Cimento {\bf A112}, 1469 (1999)
\bibitem{ref10} J.~P.~Ostriker, astro-ph/9912548
\bibitem{ref11} D.~N.~Spergel and P.~J.~Steinhardt, Phys.~Rev.~Lett.~{\bf 84}, 3760 (2000)
\bibitem{ref12} N.~Yoshida et al., ApJ {\bf 535}, L103 (2000)
\bibitem{ref13} A.~Burkert, astro-ph/0012178, and references therein
\bibitem{ref14} J.~R.~Primack, astro-ph/0205391
\bibitem{ref15} M.~Duff and P.~van Nieuwenhuizen, Phys.~Lett.~{\bf 94B}, 179 (1980)\\
 M.~Duff , Phys.~Lett.~{\bf B226}, 36 (1989)
\bibitem{ref16} M.~Turok and S.~W.~Hawking, hep-th/9803156v2
\end{thebibliography}
\end{document}